\documentclass[11pt,twoside]{article}
\usepackage{asp2010}

\resetcounters

\begin{document}

\title{A remarkable sample of new symbiotic stars towards the Galactic Bulge}
\author{Brent Miszalski,$^{1,2}$ Joanna Miko{\l}ajewska$^{3}$ and Andrzej Udalski$^{4}$ 
\affil{$^1$South African Astronomical Observatory, PO Box 9, Observatory, 7935, South Africa}
\affil{$^2$Southern African Large Telescope Foundation, PO Box 9, Observatory, 7935, South Africa}
\affil{$^3$Nicolaus Copernicus Astronomical Centre, Bartycka 18, 00716 Warsaw, Poland}
\affil{$^4$Warsaw University Observatory, Al. Ujazdowskie 4, PL-00-478, Warsaw, Poland}}

\begin{abstract}
   Symbiotic stars are the longest orbital period interacting binaries, where nova-like outbursts are generated by the accretion of a high mass loss rate red giant wind onto a white dwarf companion. Long-term photometric monitoring surveys such as OGLE and MACHO are ideal platforms to identify nova-like events in symbiotic stars, however there are only a handful of known systems within the small footprint of these surveys. We introduce a systematic Halpha emission line object survey for new symbiotic stars covering 35 deg$^2$ towards the Galactic Bulge that combines deep 2dF/AAOmega spectroscopy with OGLE and MACHO photometry. This powerful combination has uncovered nearly two dozen new symbiotic stars, more than a dozen probable symbiotic stars, and several other unusual H$\alpha$ emission line stars. While we don't find any nova-like activity, the lightcurves do exhibit semi-regular and Mira pulsations, orbital variations and slower changes due to dust. Here we introduce a few of the new symbiotics, including H1-45, only the fourth known carbon symbiotic Mira. This remarkable discovery may be the first luminous carbon star belonging to the Galactic Bulge, according to its period-luminosity relation distance of $6.2\pm1.4$ kpc, potentially shedding new light on the puzzling lack of luminous carbon stars in the Bulge. We also present two old novae captured in the nebular phase, complementing other surveys to better characterise the old nova population.
\end{abstract}

\section{Introduction}
There may be as many as $10^4$--$10^5$ Galactic symbiotic stars \citep{MR92,K93}, however observational campaigns have amounted to no more than $\sim$300 known \citep{Allen84,Kenyon86,Mik97,Bel00,Corradi08,Corradi10}. We are motivated to find more symbiotics to find potential SN Ia progenitors (e.g. RS Oph) and to better compare their birth rate to SN Ia rates. \citet{Corradi12} and ref. therein have demonstrated the power of H$\alpha$ surveys to discover more symbiotics and identified the Galactic Bulge as a crucial environment to address the latter problem. A preliminary sample of new Bulge symbiotics was identified by \citet{misz09a} after a systematic spectroscopic survey of H$\alpha$ emission line candidates covering 35 deg$^2$ towards the Bulge. Here we introduce some results from a comprehensive reanalysis of the survey which has identified several new symbiotic stars (Sect. \ref{sec:obs}), as well as a wide variety of other unusual H$\alpha$ emitters, amongst which we found two old novae (Sect. \ref{sec:novae}). The symbiotic-specific Raman scattered OVI emission bands were found in 35\% of the nearly two dozen new symbiotics. Recovery of 11/13 known symbiotics from \citet{Bel00} and other H$\alpha$ emitters from \citet{KW03} testify to the high completeness of the survey. We refer the reader to \citet*{mmu13} for full details.

\section{Representative observations}
\label{sec:obs}
\begin{figure}[h]
   \begin{center}
      \includegraphics[scale=0.45,angle=270]{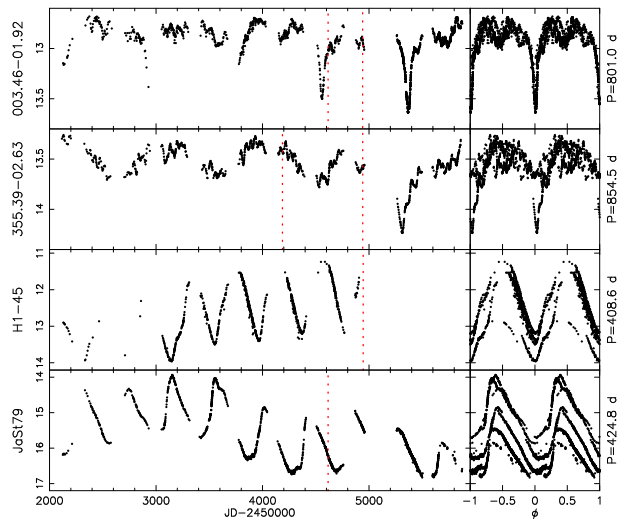}
   \end{center}
   \caption{OGLE lightcurves of two S-types (003.46-01.92 and 355.39-02.63) and two D-types (H1-45 and JaSt79). The dotted red vertical lines mark epochs when spectra were taken with 2dF/AAOmega.}
   \label{fig:lcs}
\end{figure}

Crucial to the success of our survey was the combination of deep AAT 2dF/AAOmega spectroscopy \citep{Lewis02,Sharp06}, covering 3700--8850 \AA\ at 3.5--5.3 \AA\ resolution, and long-term lightcurves from the OGLE \citep{Udalski08} and MACHO \citep{Alcock97} projects. These lightcurves reveal semi-regular or Mira pulsations, orbital variations or slow variations due to dust. Figure \ref{fig:lcs} shows the periodic lightcurves of two new S-types and two new D-types. \citet{Gromadzki09} and \citet{Lutz2010} have found similar variability amongst previously known symbiotics. Such variability is particularly useful to identify symbiotics previously misclassified as PNe and was used to determine orbital periods for 5 S-types and Mira pulsation periods for 3 D-types. The lightcurve of the eclipsing 003.46-01.92 also shows ellipsoidal variations, which makes this object of interest to monitor for nova-like activity \citep{Mik13}. Unfortunately, we noticed no nova-like outbursts amongst our sample, suggesting they are rare over timescales of $\sim$7--15 years.

\begin{figure}
   \begin{center}
      \includegraphics[scale=0.25,angle=270]{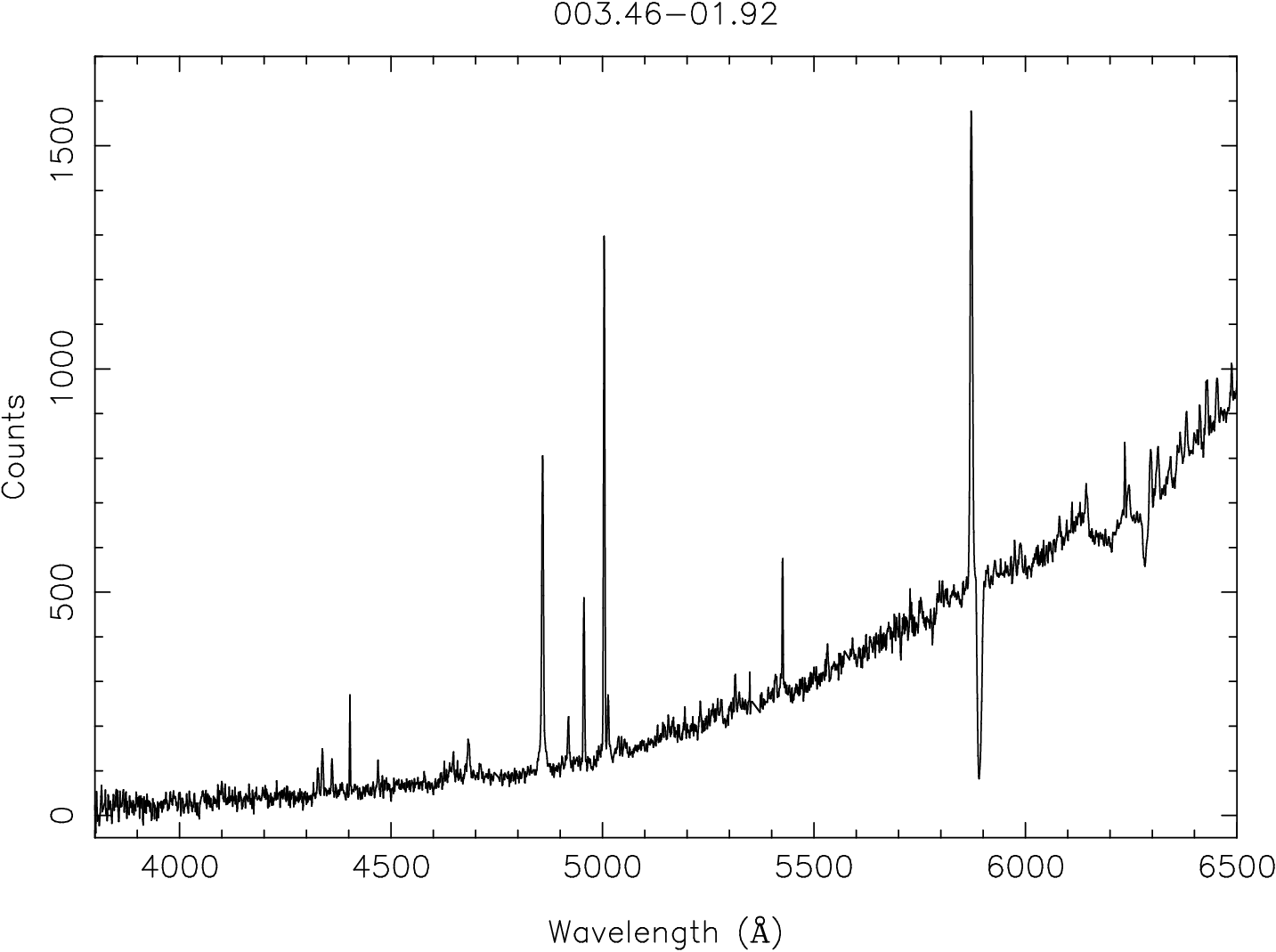}
      \includegraphics[scale=0.25,angle=270]{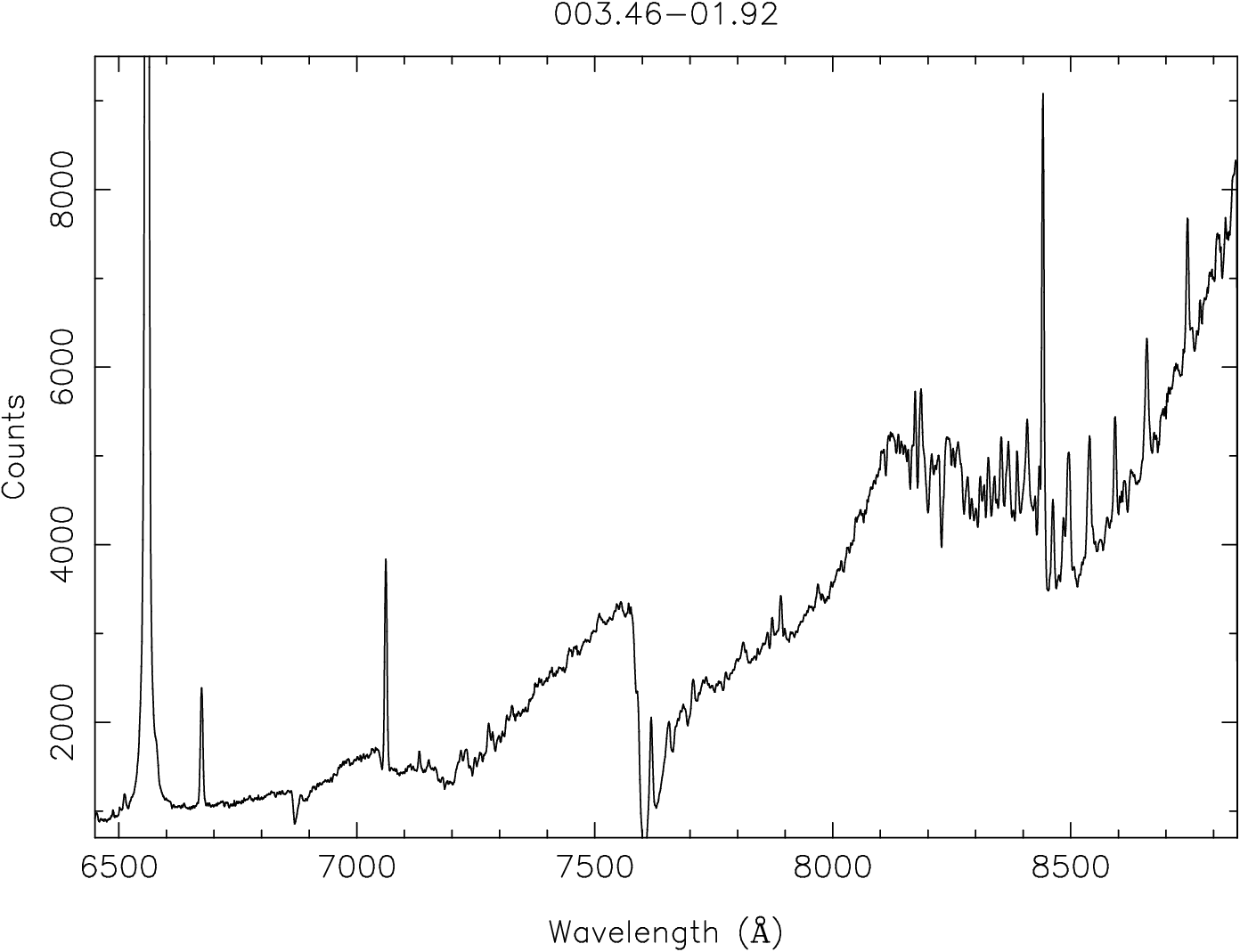}
      \includegraphics[scale=0.25,angle=270]{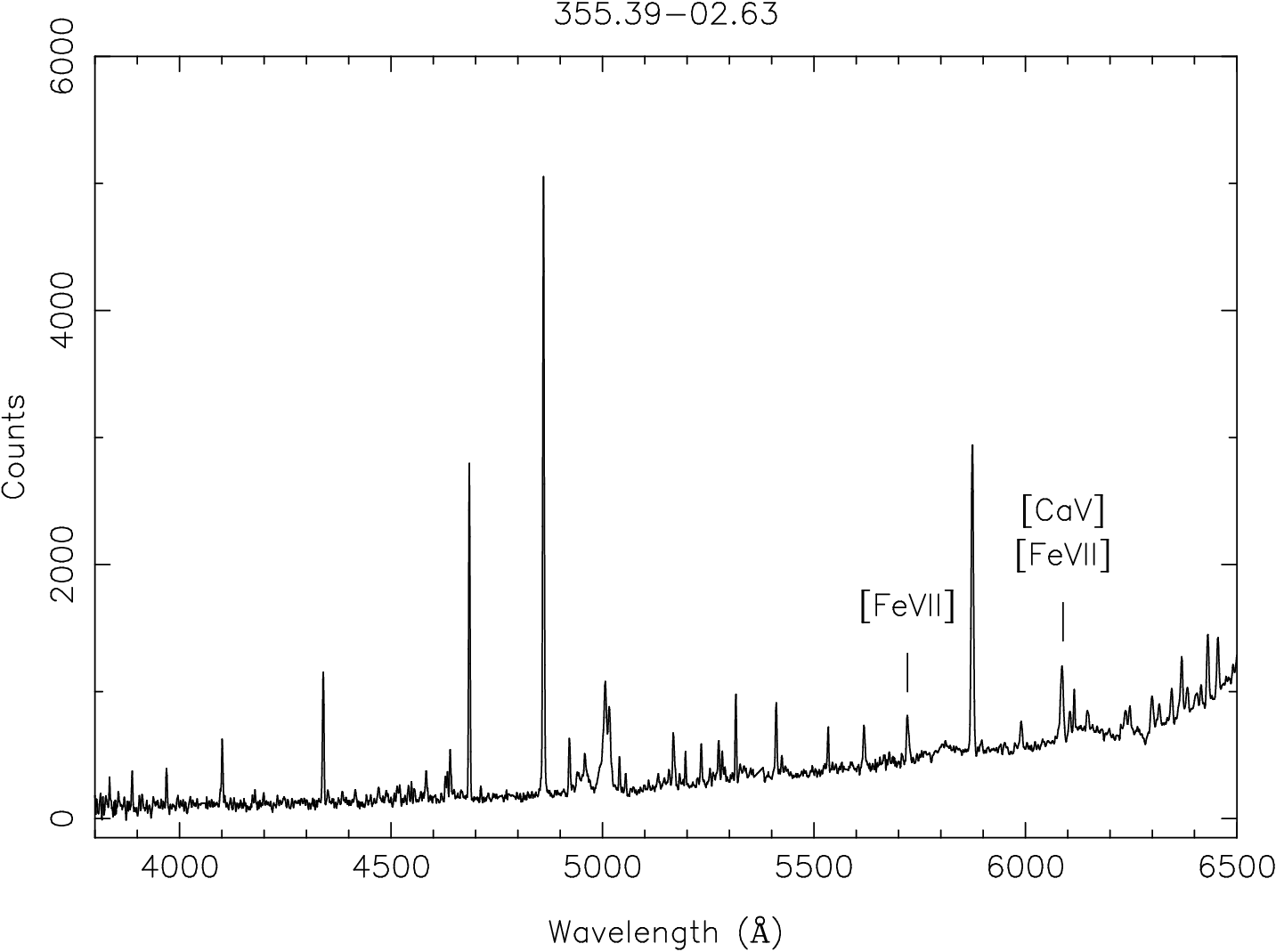}
      \includegraphics[scale=0.25,angle=270]{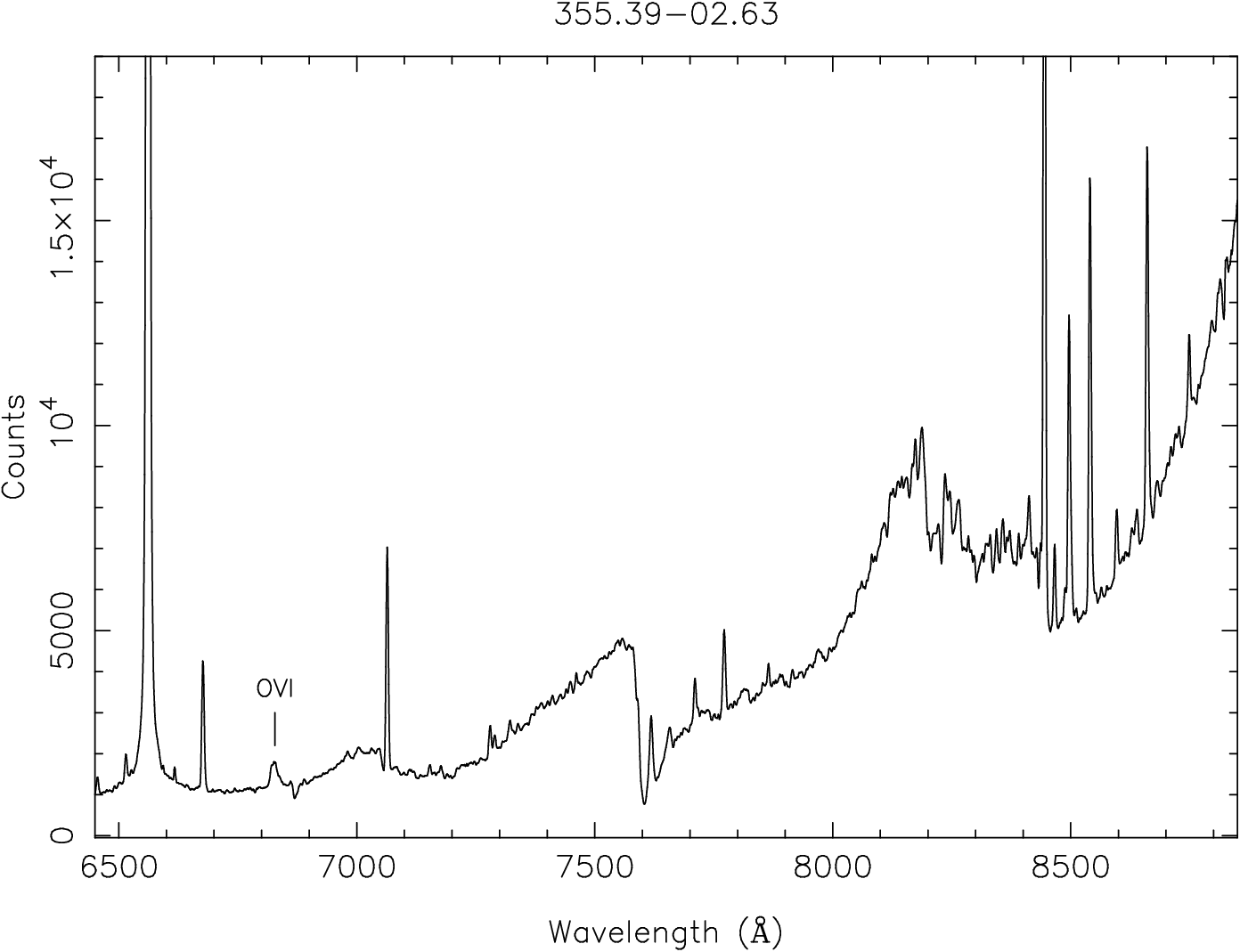}
      \includegraphics[scale=0.25,angle=270]{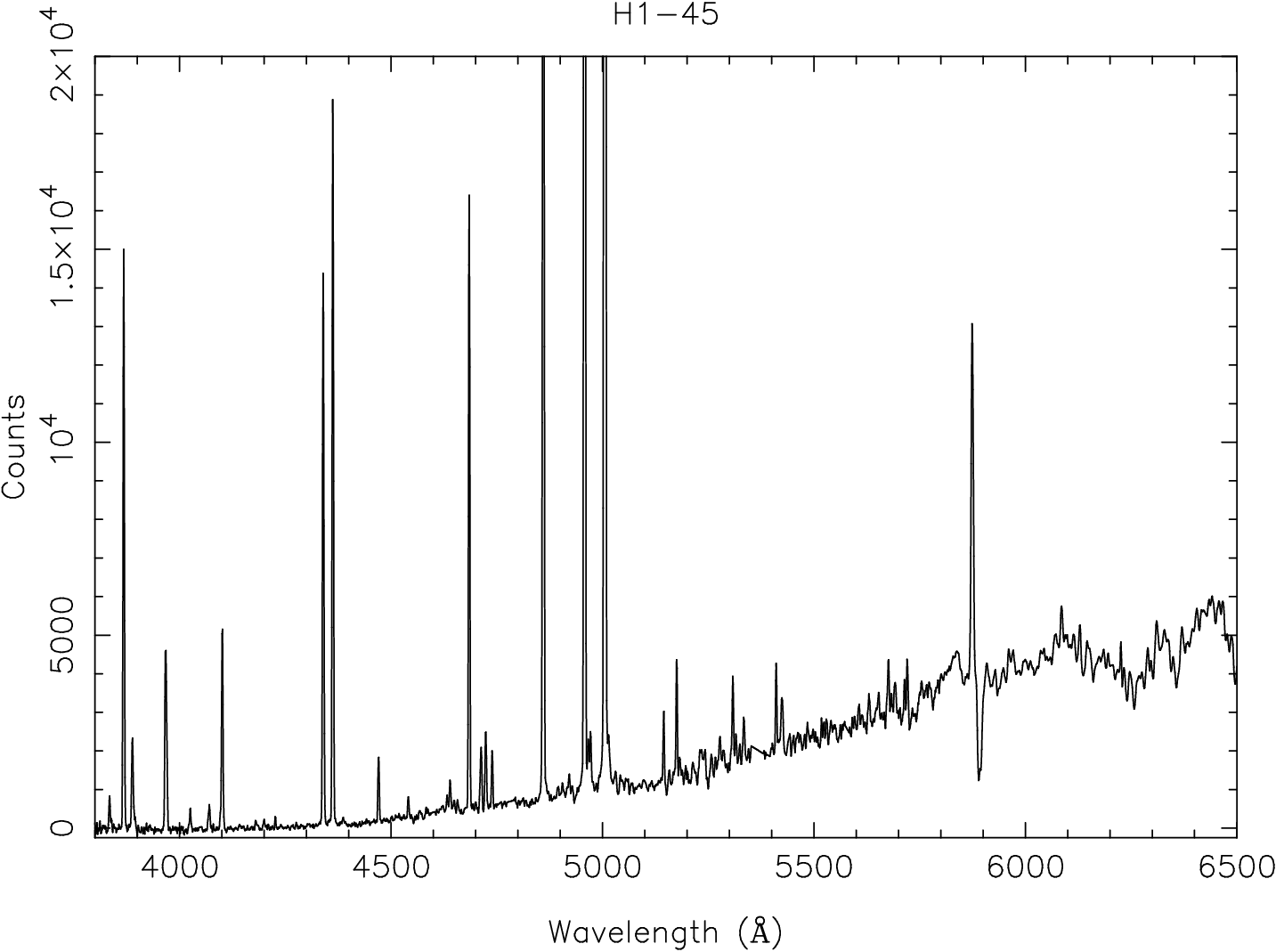}
      \includegraphics[scale=0.25,angle=270]{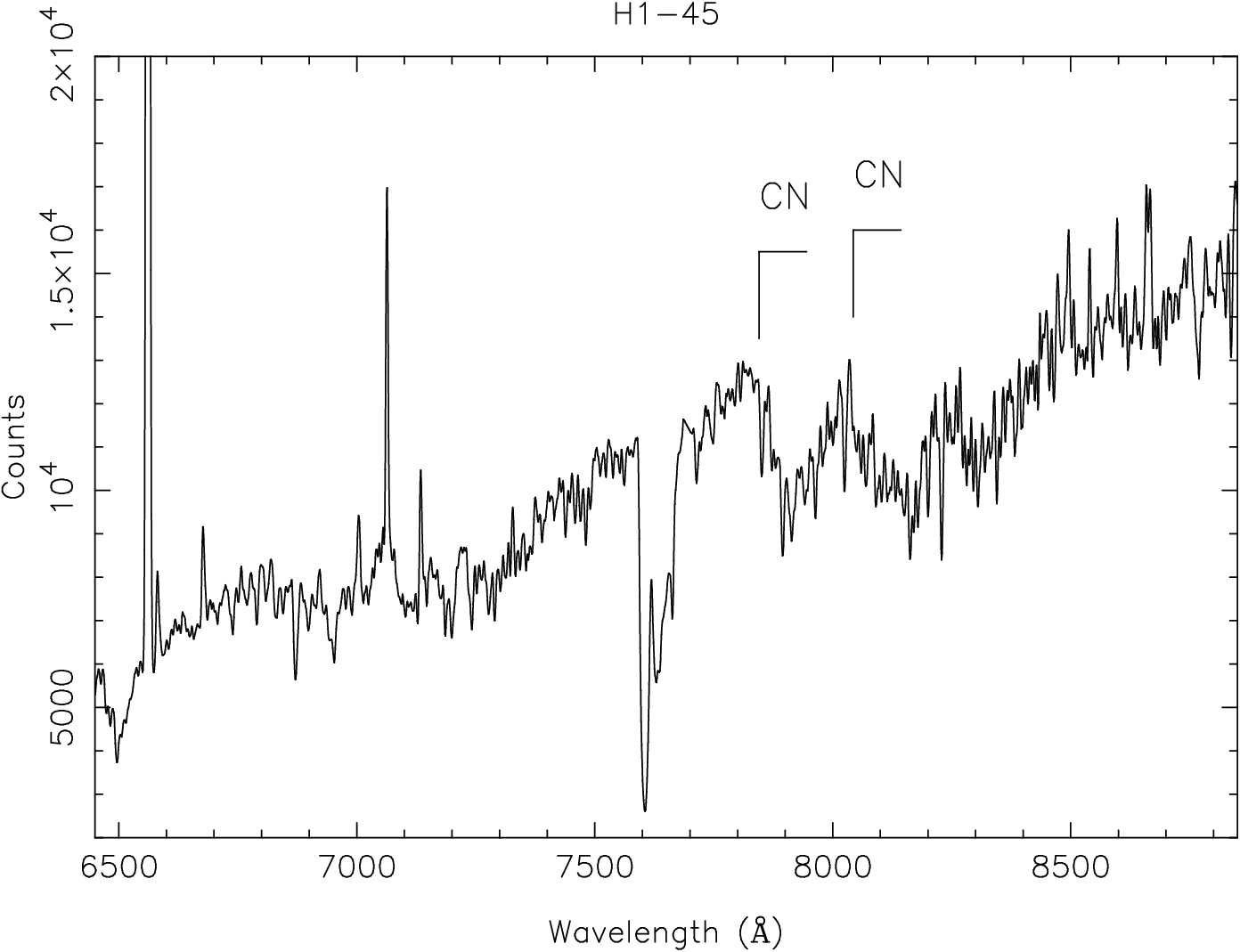}
      \includegraphics[scale=0.25,angle=270]{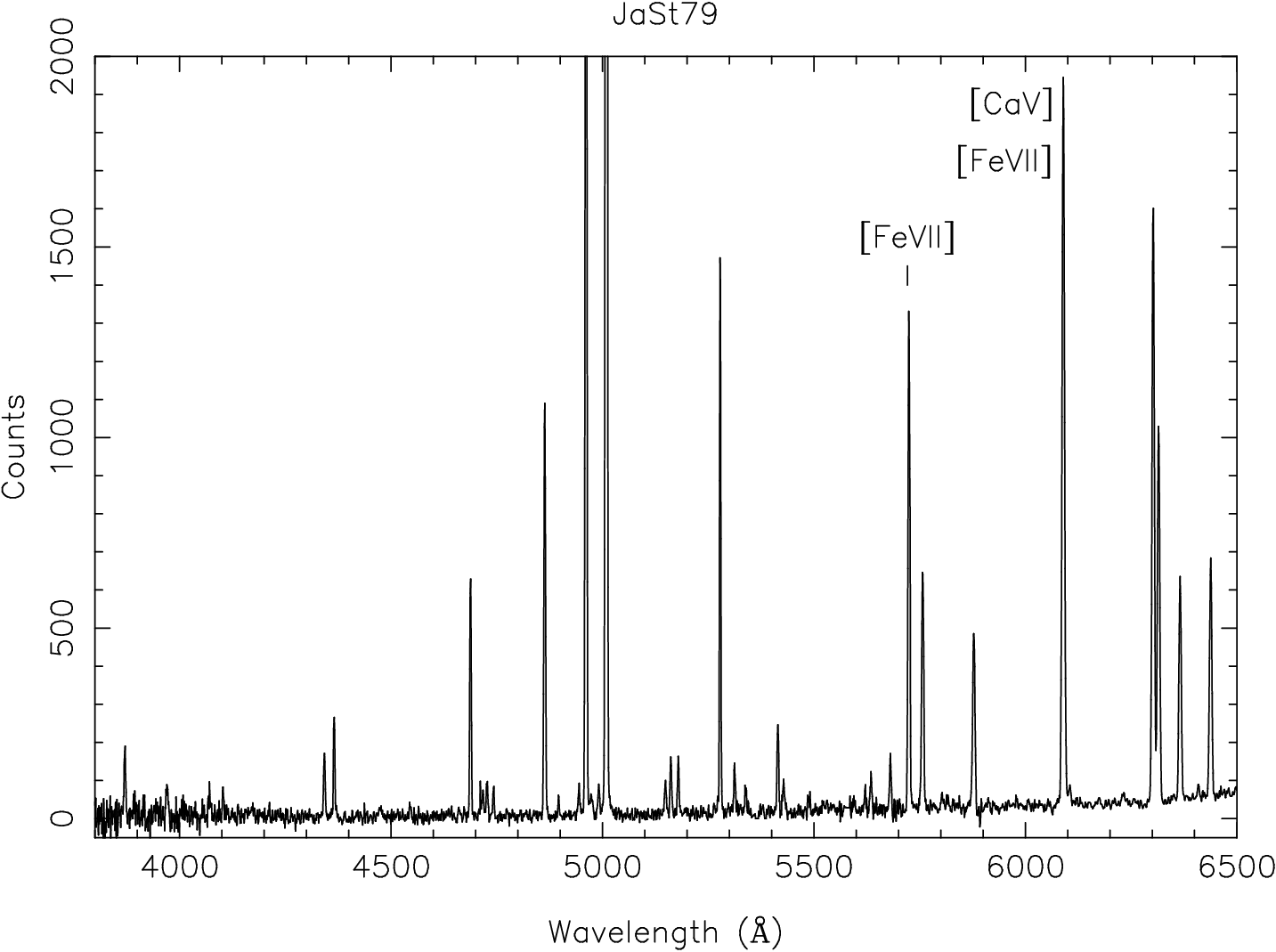}
      \includegraphics[scale=0.25,angle=270]{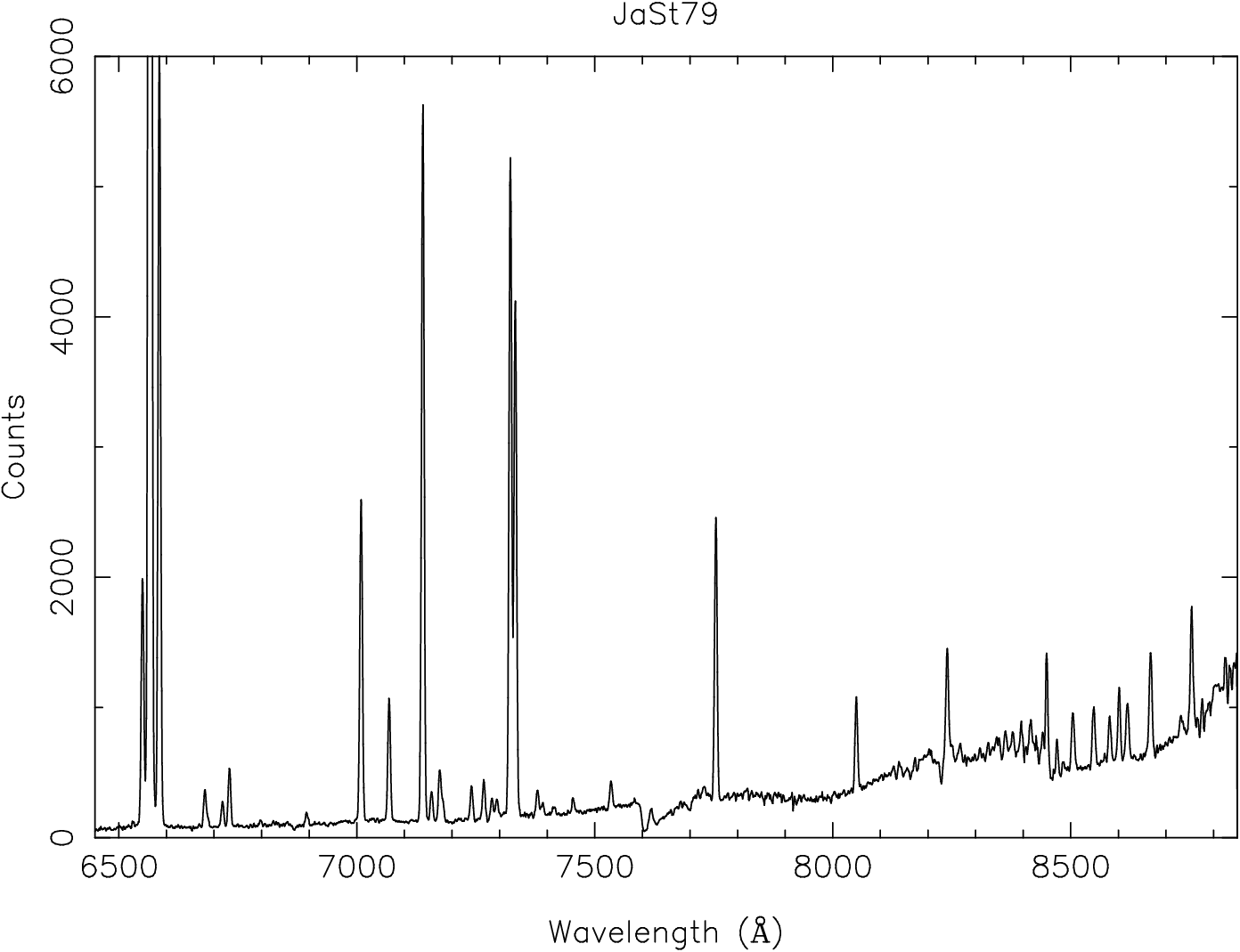}
   \end{center}
   \caption{The 2dF/AAOmega spectra of the symbiotics in Fig. \ref{fig:lcs}.}
   \label{fig:syspec}
\end{figure}

Spectra of the new symbiotics in Fig. \ref{fig:lcs} are shown in Fig. \ref{fig:syspec}. In the case of H1-45 and JaSt79 the symbiotic Miras \citep{Whitelock87} are not completely extinguished by their dusty coccoons and are visible in the AAOmega spectra. This is not always the case and makes it harder to confirm D-type symbiotics using only optical observations, explaining the dominance of D-types amongst our list of probable symbiotics. As demonstrated by RX Pup, only near-infrared observations can reveal Mira pulsations in the most heavily obscured systems \citep{Mik99}. In the probable D-types of our sample slow variations due to dust are common. 

The strong CN bands seen in H1-45 are remarkable, making H1-45 only the fourth known Galactic symbiotic carbon Mira \citep{Gromadzki09,Corradi11}. Furthermore, its period-luminosity distance of $6.2\pm1.4$ kpc using the \citet{Whitelock08} relation suggests H1-45 may be the first luminous carbon Mira in the Galactic Bulge. At its Galactic coordinates ($\ell,b$)=(2.0,$-$2.0) it may reside on the near-side of the inner bar \citep{Gonzalez11}. This would be hard to reconcile with the lack of carbon Miras in the Bulge, a longstanding unsolved problem (see e.g. \citet{Feast07}), unless the carbon enhancement was accreted from the progenitor of the white dwarf companion.

\section{A couple of old novae}
\label{sec:novae}
\begin{figure}
   \begin{center}
      \includegraphics[scale=0.25,angle=270]{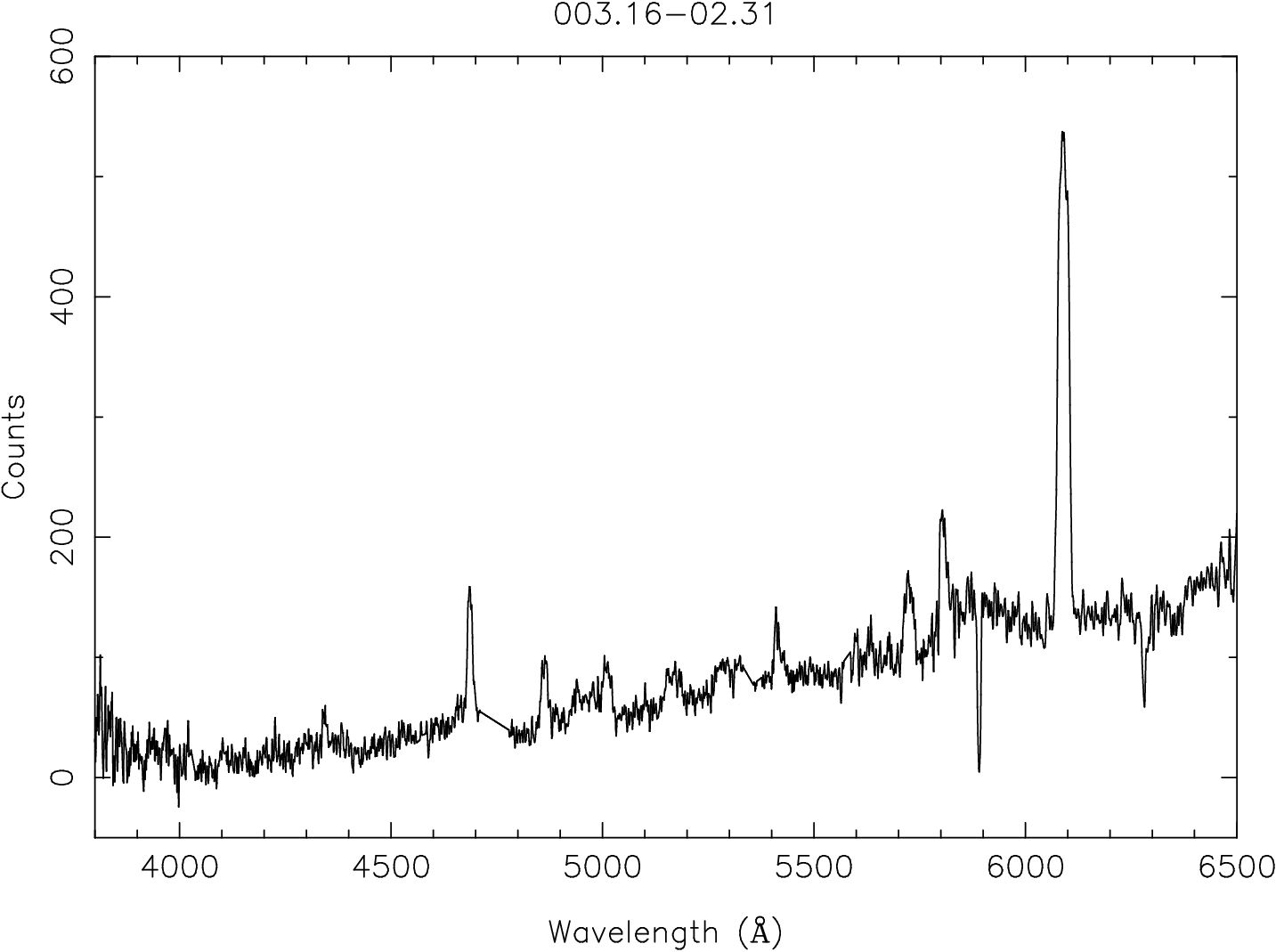}
      \includegraphics[scale=0.25,angle=270]{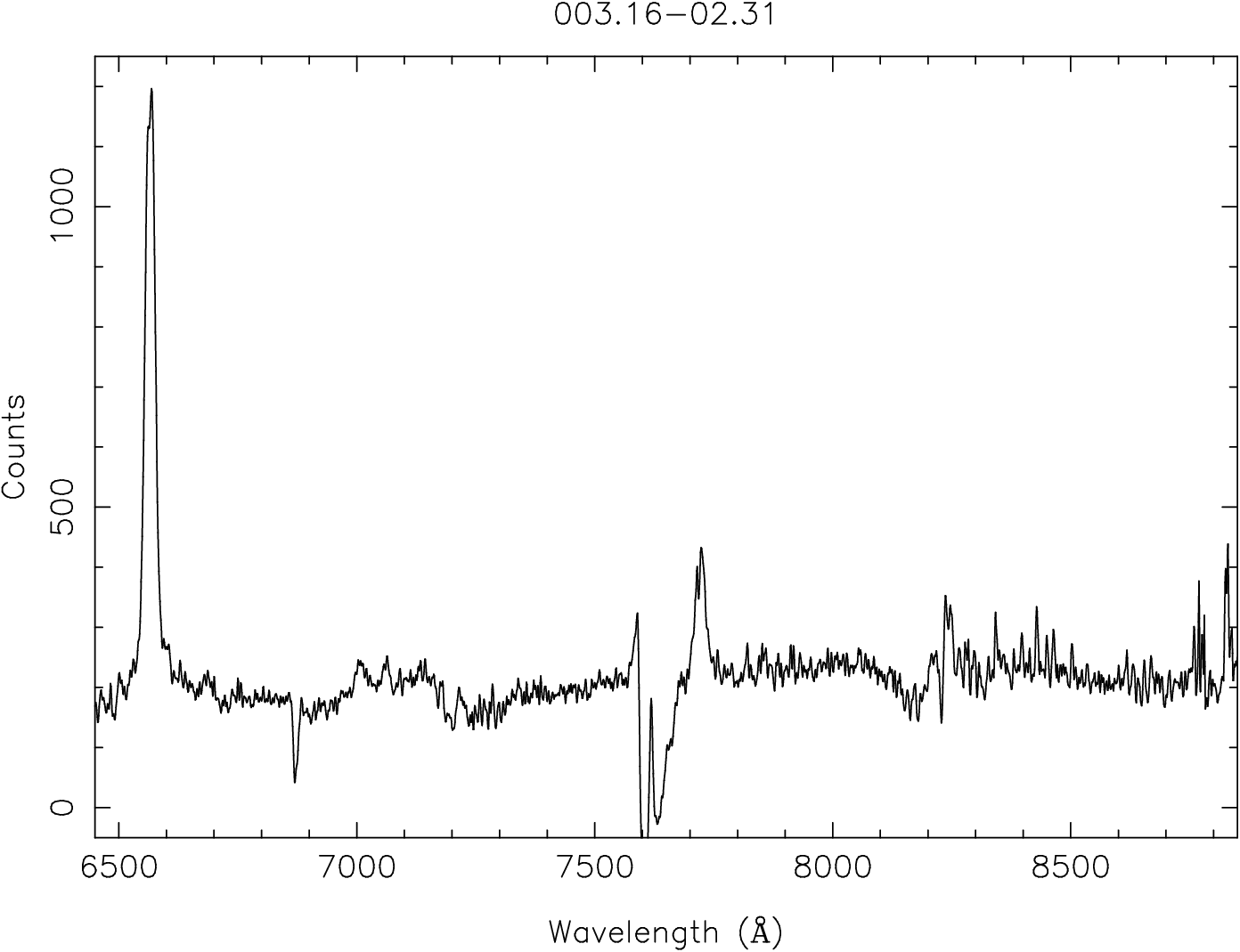}
      \includegraphics[scale=0.25,angle=270]{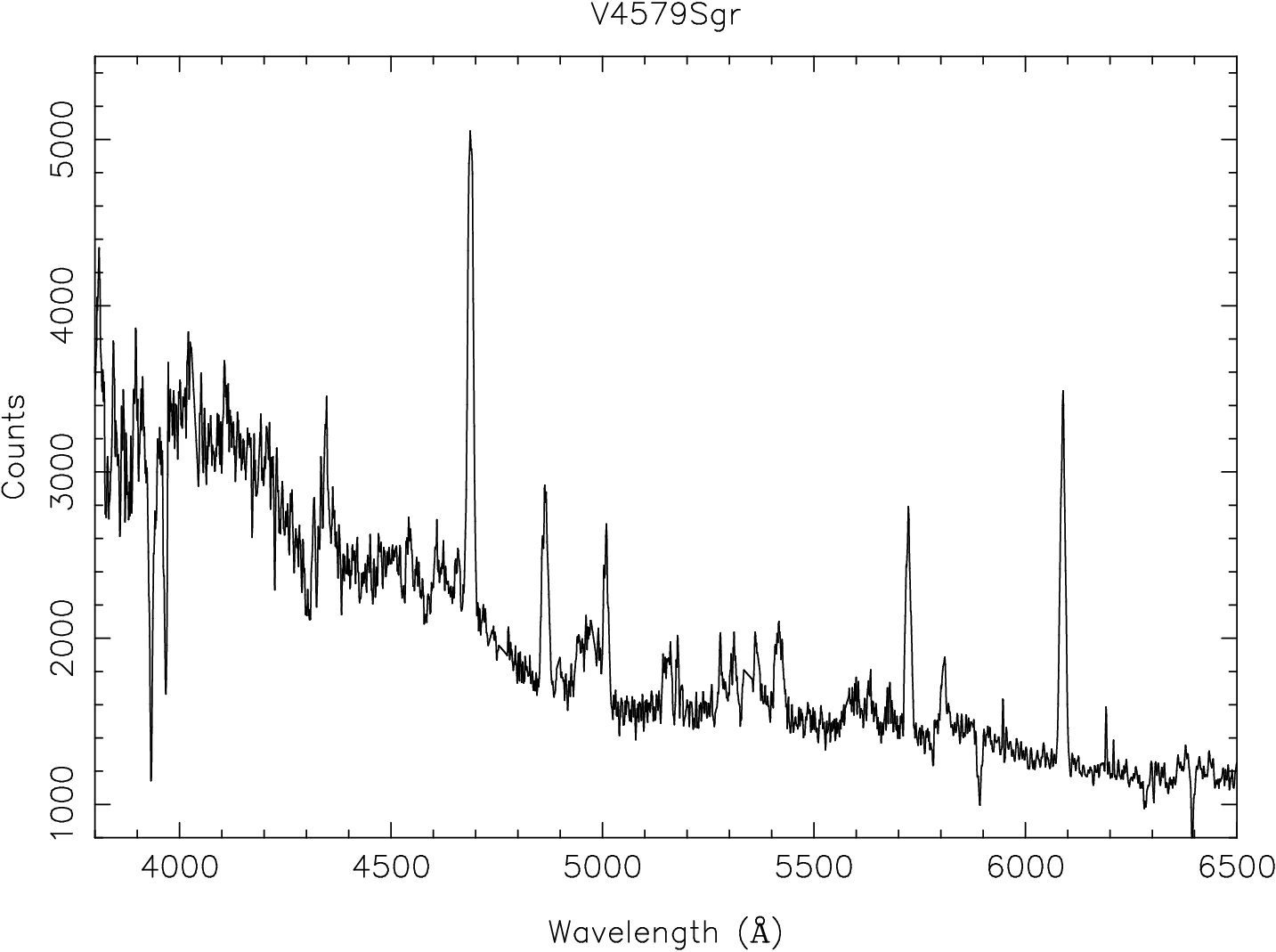}
      \includegraphics[scale=0.25,angle=270]{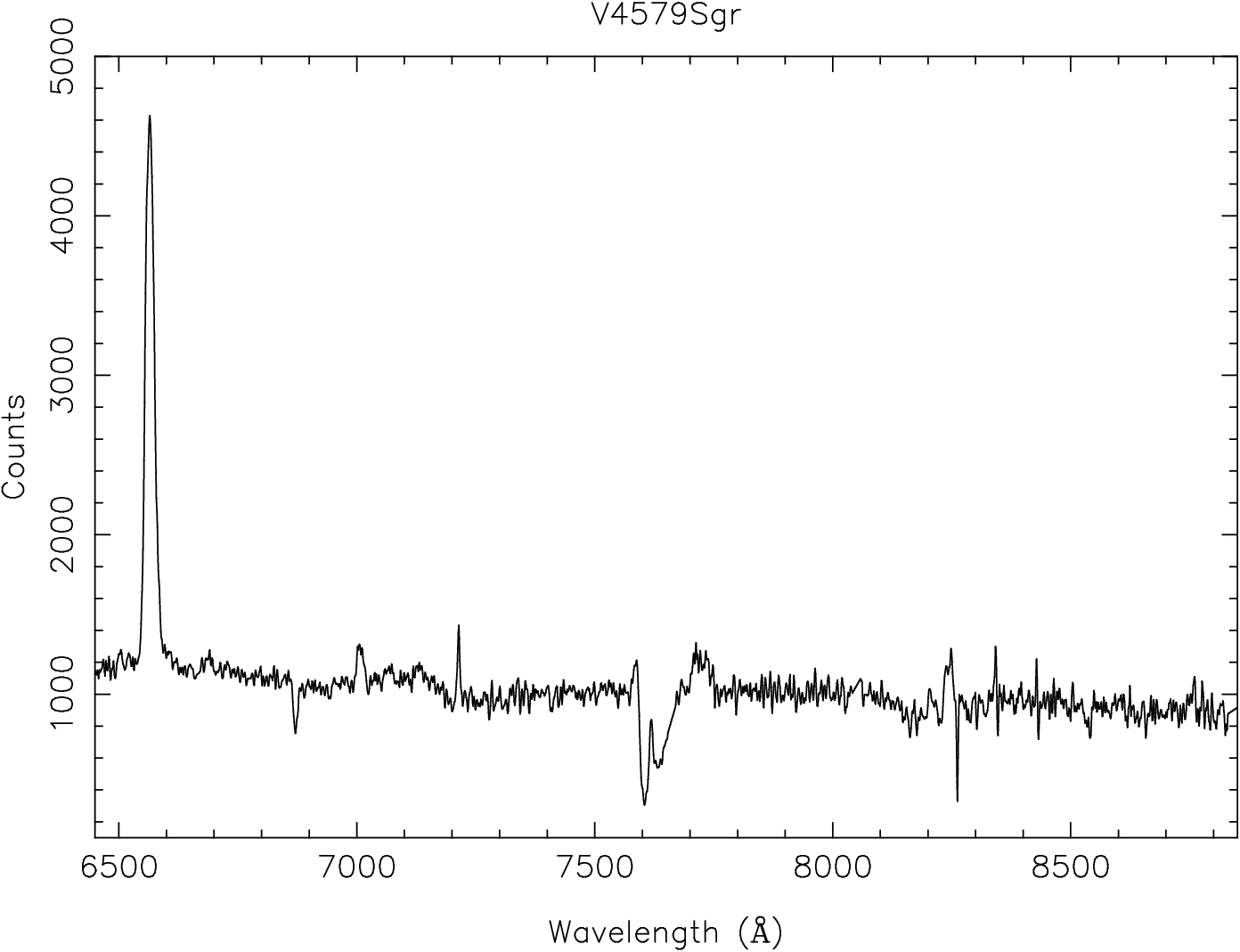}
   \end{center}
   \caption{The 2dF/AAOmega spectra of old novae 003.16-02.31 (top) and V4579 Sgr (bottom).}
   \label{fig:novaspec}
\end{figure}

\begin{figure}
   \begin{center}
      \includegraphics[scale=0.25,angle=270]{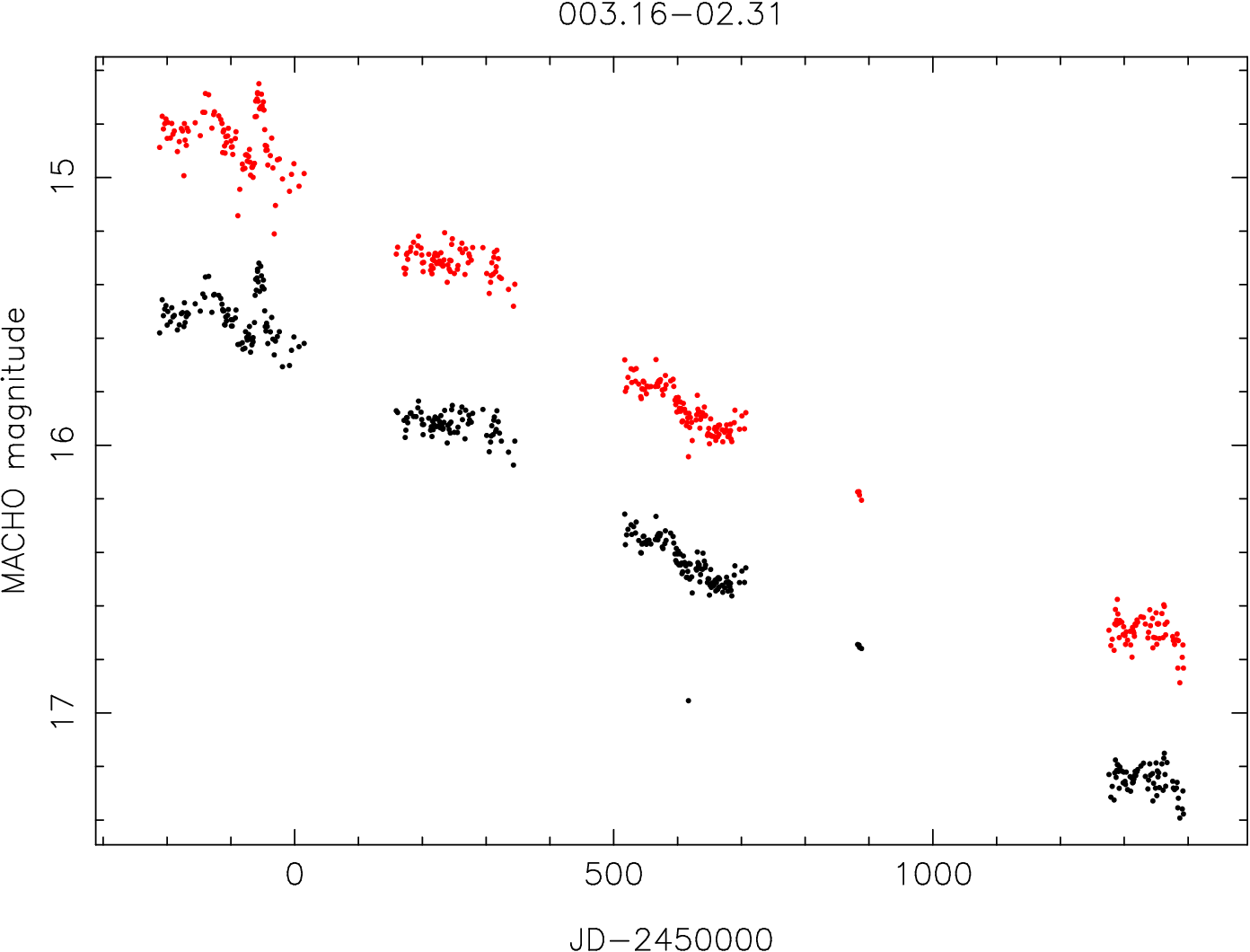}
      \includegraphics[scale=0.25,angle=270]{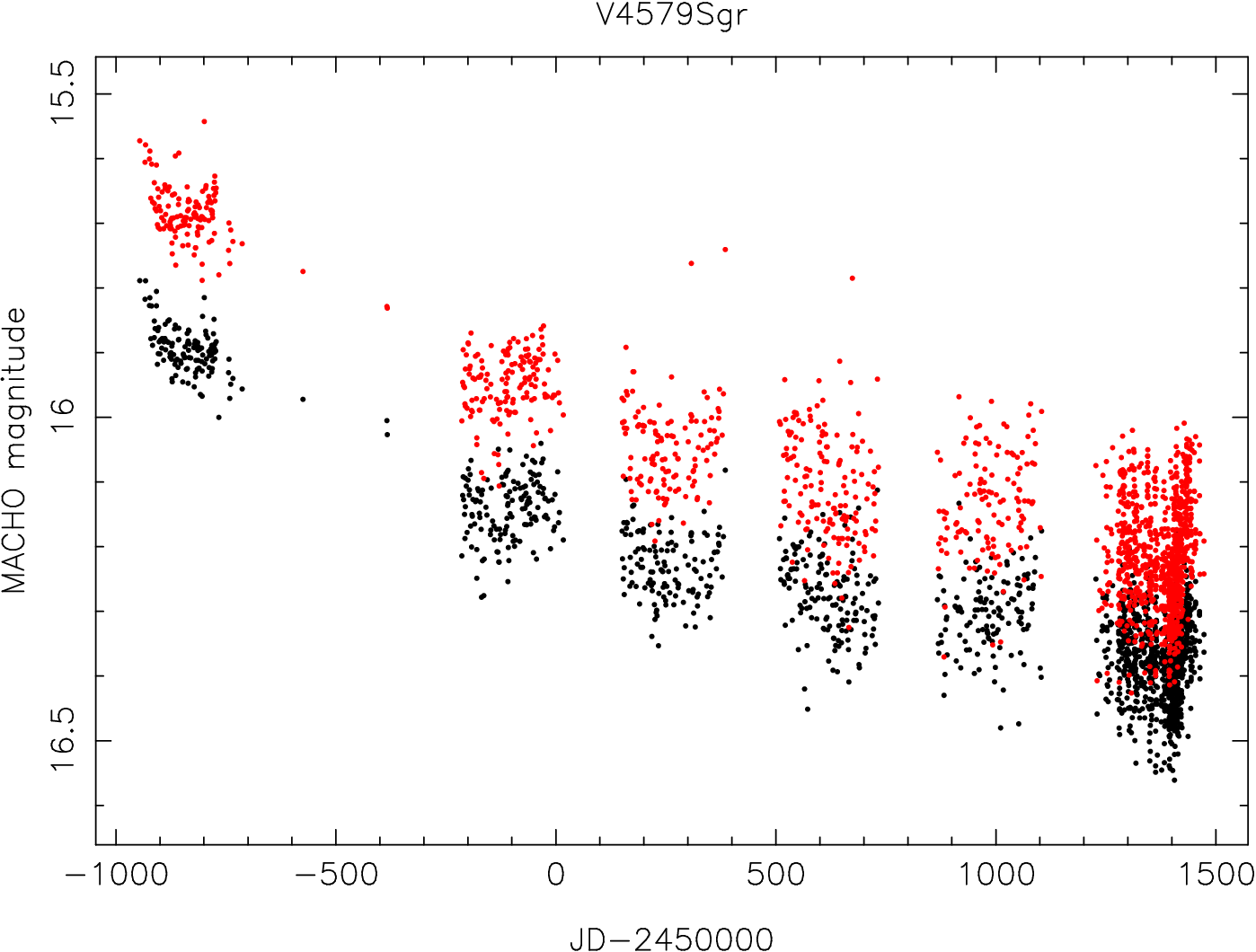}
   \end{center}
   \caption{MACHO lightcurves of the two old novae in $V$ (black) and $R$ (red) bands transformed using the relations given by \citet{Lutz2010}.}
   \label{fig:novalc}
\end{figure}

Amongst the various other H$\alpha$ emitters in our survey were two novae, 003.16-02.31 and V4579 Sgr. Their spectra are typical of He/N novae in the nebular phase (Fig. \ref{fig:novaspec}) and the MACHO lightcurves beautifully record their decline (Fig. \ref{fig:novalc}). Unfortunately, the lightcurve maximum of each nova was not covered by MACHO, but we measured the decline rates to be $1.2\times10^{-3}$ and $1.9\times10^{-4}$ mag day$^{-1}$, respectively. The OGLE lightcurves are considerably flatter \citep{mmu13}. \citet{Woudt05} found an orbital period of 0.117 days in the eclipsing lightcurve of 003.16-02.31, the same object as their `Sgr (MACHO peculiar variable)'. These discoveries contribute to other programmes that aim to better characterise the old nova population \citep{Walter12,Tappert12}.

\section{Future work}
Apart from extending this survey, we have commenced similar work in the Magellanic Clouds where there are only a handful of symbiotics known \citep{Bel00}. In particular, we are currently analysing SALT spectroscopy for the symbiotic star candidates identified by \citet{misz2011}, as well as several Large Magellanic Cloud symbiotic stars. Results from this spectroscopic survey will be reported elsewhere.

\section{Acknowledgements}
We thank the conference organisers, in particular Patrick and Valerio, for a great conference and the opportunity to present this work.
   This work benefited from the framework of the European Associated Laboratory ``Astrophysics Poland-France''.
JM is supported by the Polish National Science Center grant number DEC-2011/01/B/ST9/06145.
   The OGLE project has received funding from the European Research Council under the European Community's Seventh Framework Programme (FP7/2007-2013)/ERC grant agreement No. 246678.
\bibliography{sy}
\end{document}